\documentclass[intlimits,twoside,a4paper]{article}

\usepackage{amsmath,amssymb}
\usepackage{graphicx}
\usepackage{color}

\usepackage[T2A]{fontenc}
\usepackage[cp1251]{inputenc}

\usepackage[eqsecnum]{cmpj2}

\issue{2015}{18}{4}{43001}
\doinumber{10.5488/CMP.18.43001}

\title[First sound velocity in liquid $^4$He]%
{First sound velocity in liquid $^4$He%
}
\author[O.I. Hryhorchak]{O.I. Hryhorchak}
\address{Department for Theoretical Physics, Ivan Franko National
University of Lviv, \\ 12 Drahomanov St., 79005 Lviv, Ukraine}

\date{Received April 1, 2015, in final form June 25, 2015}
\authorcopyright{O.I. Hryhorchak, 2015}

\begin{document}

\maketitle

\begin{abstract}
Based on the many-boson system structure factor, which takes into account three-
and four-particle direct correlations, there was found the first sound velocity temperature behaviour
in liquid $^4$He in the post-RPA approximation.
The expression received  for the sound velocity matches with the well-known results in both low and high
temperature limits. The results of this paper can be used to analyze the
contributions of three- and four-particle correlations into thermodynamic and structural features
of  liquid $^4$He.
\pacs 05.30.Jp, 67.25.-k, 43.35.+d, 62.60.+v,67.25.dt
\end{abstract}

\section{Introduction}

The study of the sound velocity in Fermi and Bose systems has been attracting attention of researchers for a long time and
still remains relevant today \cite{WON,OzSt,JCLK,MPZ,KMKT,LiHu,STSM,MAOK,NYAS}. There are a lot of various
many-particle systems,
in which these investigations take place nowadays, for example: mixtures of $^3$He and $^4$He \cite{LNBM,SCKR,ZKRU},
$^4$He in aerogel \cite{MAOK,NYAS}, solid $^3$He \cite{STSM}, plasma \cite{Stol}, trapped Bose gas \cite{LiHu}, etc.

Experimental study of the sound velocity temperature behavior in many-boson system began almost 80 years ago.
Barton \cite{Burton} and a year later Findlay with colleagues \cite{Findlay,Findlay2} using ultrasonic method
made measurements in liquid $^4$He under saturated vapor pressure (experiments were also
carried out at higher pressures). The range of the research began from temperature $T = 1.75$~K.
The precision of measurements was estimated less than 0.5 percent by authors.
The curve of the then obtained temperature behavior  perfectly conforms with the current results \cite{DonBar}.

Hronevold \cite{Groenewold} interpreted these experimental data.
He proved that the ultrasonic waves in helium II are adiabatic despite their high frequency and
the large thermal conductivity of helium. He also made attempts to explain
a much smaller gap of the sound velocity at the $\lambda$-point transition compared to the value
which stems from the Ehrenfest equation.

Another issue that caused interest of researchers was the sound attenuation in the $^4$He liquid.
Pellet and Skvayyer \cite{Pellam} sought it experimentally
and showed that the experimental data are in a good agreement (at least for the temperature region from $3.2$~K to $4.2$~K)
with theoretical results obtained from the classical formulae
in assumption that the reasons of sound attenuation are only viscosity and heat losses.
Later on, the attenuation and the sound velocity in $^4$He were  calculated using the Landau-Khalatnikov kinetic
equations and the phonon Boltzmann equation \cite{Maris}. In paper \cite{FoMo}, the authors compared the theoretical results
of the critical  behavior of the sound propagation in liquid $^4$He (as well as in other liquids)
near the gas-liquid critical point which was derived within the field-theoretic renormalization group formalism
with the experimental data. Another phenomenon that is studied nowadays is the first
sound reflection in liquid $^4$He \cite{Meln}.

Theoretical study of the sound velocity was also carried out by the collective variables representative method, but only in the limit of low temperatures
\cite{UhnVak77,VakUhn79,VHU79,VakUhn80,Vak1990_2}.
Particularly in papers \cite{VakUhn79,VakUhn80}, correlation was found between
the first sound velocity at absolute zero temperature and the Fourier coefficient of pair interparticle interaction energy
in the ``one sum over the wave vector'' approximation. This result was obtained in two different ways.
The first one went through finding
longwave asymptotics of structural quantities, which take into account direct three- and four-particle correlations,
and the second one was based
on the well-known thermodynamic relation: $mc^2/N=\partial^2E_0/\partial N^2$, where $c$ is the first sound velocity,
$N$ is the number of particles, $m$ is the particle mass, $E_0 $ is ground state energy
taken in the pair correlations approximation. The above mentioned relations in higher approximation
(``two sums over the wave vector'') were found in paper \cite{VakHryh4}.

 The objective of this paper is to find the temperature behavior of the first sound velocity
 in liquid $^4$He. We proceed from the exact relation
 that links the sound velocity with the longwave limit of the structure factor \cite{Vak2004}, as well as from
 the results which we obtained earlier for the pair
 structure factor in the ``one sum over the wave vector'' approximation that takes into account three- and four-particles
 correlations \cite{VakHryh3}. It should be noted that the calculation of the sound velocity through the structure factor in
 pair correlations approximation only leads to a constant value in pre-critical region and gives a light growth
 in the above-critical region (this behavior does not conform with the experimental data).
 The results of the sound velocity calculation in the post-RPA approximation show
 a fairly good agreement with the experimental data. The resulting expression for the first sound velocity
 in the limit of low temperatures
 matches with the already known one \cite{VakUhn79,VakUhn80}.

 \section{Structure factor and first sound velocity in many-boson system}

It is known \cite{Vak2004} that there is a relation between the sound velocity value in the wide temperature region and
the longwave limit of the structure factor:
\begin{align}\label{Sc}
\lim\limits_{q\rightarrow 0}S(q)=\frac{T}{mc^2(T)}\,.
\end{align}
Sound velocity temperature behaviour obtained from the expression for the structure factor in
 pair correlations approximation yields incorrect results in the above-critical region,
 as it was already mentioned. Therefore, we
should use the post-RPA approximation
which takes into account three- and four-particle correlations \cite{VakHryh4}:
\begin{align}
 S_q=\frac{S_0(q)}{1+(\lambda_{q}+\Pi_q)S_0(q)}\,,
\end{align}
where $S_0(q)$ is the two-particle structure factor of the ideal bose gas,
\begin{align}\label{S2}
\Pi_q =&-
\frac{1}{2N}\frac{1}{S_0^2(q)}\sum_{\mathbf{k}\neq0}\frac{\lambda_{k}S_0^{(4)}({\bf q},-{\bf q}, {\bf k},-{\bf
k})}{1+\lambda_{k}S_0(k)}
+\frac{1}{2N}\frac{1}{S_0^2(q)}\sum_{\mathbf{k}\neq0}
\frac{\lambda_k\lambda_{|{\bf q}+{\bf k}|}\big[S_0^{(3)}({\bf q},{\bf k}, -{\bf
q}-{\bf k})\big]^2}{[1+\lambda_{k}S_0(k)][1+\lambda_{|{\bf q}+{\bf k}|}S_0(|{\bf q}+{\bf k}|)]}
\nonumber\\
&+ 4C_2({\mathbf{q}})+\frac{12}{N}\sum_{\mathbf{k}\neq0}\frac{C_3({\mathbf
q},{\mathbf k}, -{\bf q}-{\mathbf k})S_0^{(3)}({\bf q},{\bf
k},-{\bf q}-{\bf k})}{S_0(q)[1+\lambda_{k}S_0(k)][1+\lambda_{|{\bf q}+{\bf k}|}S_0(|{\bf q}+{\bf k}|)]}
+\frac{8}{N}\sum_{\mathbf{k}\neq0}\frac{C_4({\mathbf q},{\mathbf
k})S_0(k)}{1+\lambda_{k}S_0(k)}\nonumber\\
&+\frac{72}{N}\sum_{\mathbf{k}\neq0}
  \frac{C_3^2({\mathbf q},{\mathbf k},-{\bf q}-{\bf k})S_0(k)S_0(|{\bf q}+{\bf k}|)}{[1+\lambda_{k}S_0(k)][1+\lambda_{|{\bf q}+{\bf k}|}S_0(|{\bf q}+{\bf k}|)]},
\end{align}
\begin{eqnarray}\label{l_and_a}
\lambda_q=\alpha_{q}\tanh(\beta
E_q)-\tanh(\beta\varepsilon_q),
\qquad
\alpha_q=\sqrt{1+{\frac{2N}{V}\nu_q}\left\slash\frac{\hbar^2q^2}{2m}\right.}\,,
\end{eqnarray}
 $\nu_q=\int \re^{-\ri\bf{qr}}\Phi(r)\rd\bf{r}$ is the Fourier coefficient of the pair interparticle interaction energy;
 $\beta=1/T$, $T$ is temperature.

{Generally, we are not interested in the explicit form of the pair interparticle interaction energy because finally
   we express its Fourier coefficient from the experimentally measured structure factor.
   At the same time, the existence and finiteness of  the Fourier coefficient of the pair interparticle interaction
   energy in  the $^4$He liquid (it is an important issue in our theory) follow from the fact of the existence of the investigated
   system,  for which we have experimentally measured the scattering length.}

The explicit look of the expressions for the quantities $C_2({\bf q}_1)$, $C_3({\bf q}_1,{\bf q}_2,{\bf q}_3)$,
$C_4({\bf q}_1,{\bf q}_2)$ is described in paper \cite{VakHryh3}.
In the longwave limit, we obtain such expressions for these quantities as well as for $\lambda_q$:
\begin{align}
C_2^0(T)=\lim_{q\rightarrow 0}C_2({\bf q})=
\frac{1}{32N}\sum_{{\bf k}\neq0}\frac{(\alpha_{k}^2-1)^2}{\alpha_k^4}
\left[\frac{\beta^2 E_k^2}{\sinh^2\left(\beta E_k\right)}
+\beta E_k\coth\left(\beta E_k\right)-2\right],
\end{align}
\begin{align}
C_3^0(k,T)=\lim_{q\rightarrow 0}C_3({\bf q},{\bf
k},-{\bf q}-{\bf k}) &=
\frac{1}{24}\frac{\alpha_k^2+1}{\alpha_k}\tanh\left(\frac{\beta}{2}E_k\right)
-\frac{1}{12}\tanh\left(\frac{\beta}{2}\varepsilon_k\right)
+\frac{1}{48}\frac{\beta\varepsilon_k(1-\alpha_k^2)}{\cosh^2\left(\frac{\beta}{2}E_k\right)},
\end{align}
\begin{align}
C_4^0(k,T)=\lim_{q\rightarrow 0}C_4({\bf q},{\bf k})&=
\frac{1}{32}\left[
\frac{\beta^2 E_k^2(\alpha_k^2-1)^2}{4\alpha_k^3}
\frac{\tanh\left(\frac{\beta}{2}E_k\right)}
{\cosh^2\left(\frac{\beta}{2}E_k\right)}
+\frac{\beta E_k}{\cosh^2\left(\frac{\beta}{2} E_k\right)}
\frac{(\alpha_k^2-1)^2+2(\alpha_k^4-1)}{4\alpha_k^3}\right.\nonumber\\
&-\left.\frac{(\alpha_k^2-1)^2+2(\alpha_k^4-1)}{2\alpha_{k}^3}
\tanh\left(\frac{\beta}{2}E_k\right)
-\frac{4}{\alpha_k}\tanh\left(\frac{\beta}{2}E_k\right)+
4\tanh\left(\frac{\beta}{2}\varepsilon_k\right)
\vphantom{\frac{\tanh\left(\frac{\beta}{2}E_k\right)}
{\cosh^2\left(\frac{\beta}{2}E_k\right)}}\right],
\end{align}
\begin{align}
\lim_{q\rightarrow 0}\lambda_q=\beta\rho\nu_0\,,
\end{align}
where $\varepsilon_k=\hbar^2k^2/2m$, $E_k=\varepsilon_k\alpha_k$, $\rho$ is the equilibrium density of liquid $^4$He.
Using expressions for pair, three- and four-particle structure factors of ideal bose-gas (see Appendix) we can also find the
values for $1/S_0(q)$, $S_0^{(3)}({\bf q},{\bf k},-{\bf q}-{\bf k})/S_0(q)$ and $S_0^{(4)}({\bf q},{\bf k})/S_0^2(q)$
in longwave limit:
\begin{align}\label{S200}
S_2^0(T)\equiv\lim\limits_{q\rightarrow 0}
\frac{1}{S_0(k)}=
\left\{
\begin{array}{c}
0,\qquad\qquad\qquad  (T\leqslant T_\textrm{c}),
\\[1ex]
\displaystyle\frac{1}
{1+F_1(T)}, \qquad (T>T_\textrm{c}),
\end{array}\right.
\end{align}
\begin{align}\label{S300}
S_3^0(k,T)\equiv\lim\limits_{q\rightarrow 0}\frac{S_0^{(3)}({\bf q},{\bf k},-{\bf q}-{\bf k})}
{S_0(q)}=\left\{\begin{array}{c}
2n_{k}+1,\qquad\qquad\qquad\qquad\qquad (T<T_\textrm{c}),
\\[1ex]
\displaystyle  1+2\frac{S_0(k)-1+F_2(k,T)}{1+F_1(T)}, \qquad (T>T_\textrm{c}),
\end{array}\right.
\end{align}
\begin{align}\label{S400}
S_4^0(k,T)\equiv\lim\limits_{q\rightarrow 0}\frac{S_0^{(4)}({\bf q},{\bf k})}{S_0^2(q)}=
\left\{\begin{array}{c}
0,\qquad\qquad\qquad\qquad\qquad\qquad\qquad\qquad (T\leqslant T_\textrm{c}),
\\[1ex]
2[S_0(k)-1+2F_2(k,T)+F_3(k,T)], \qquad (T>T_\textrm{c}),
\end{array}\right.
\end{align}
where $T_\textrm{c}$ is the critical temperature, $n_k=[z_0^{-1}\exp(\beta\varepsilon_k)-1]^{-1}$ are occupation numbers,
$z_0=\exp(\beta\mu)$ is activity, $\mu$ is chemical potential.

The explicit look for $F_1(T)$, $F_2(k,T)$, $F_3(k,T)$ functions is as follows:
\begin{align}
 F_1(T) = \frac{1}{N}\sum\limits_{{\bf p}\neq0} n_p^2=\frac{1}{2\pi^2\rho}\int\limits_0^\infty
 \frac{p^2 \rd p}{\left[z_0^{-1}\exp(\beta p^2)-1\right]^2}\,,
\end{align}
\begin{align}
 F_2(k,T)\,= \; & \frac{1}{N}\sum\limits_{{\bf p}\neq0} n_p^2n_{|{\bf p}+{\bf k}|} =
 \frac{1}{4\pi^2\rho}\int\limits_0^\infty
 \frac{p^2 \rd p}{\left[z_0^{-1}\exp(\beta p^2)-1\right]^2}\nonumber\\
 &\times
 \left(\frac{1}{2\beta pk}
 \ln\left\{\frac{z_0^{-1}\exp\left[\beta (p+k)^2\right]-1}{z_0^{-1}\exp\left[\beta (p+k)^2\right]-1}\right\}-2\right),
\end{align}
\begin{align}
 F_3(k,T) \,= \;&\frac{1}{N}\sum\limits_{{\bf p}\neq 0}n_p^2n_{|{\bf p}+{\bf k}|}^2=
 \frac{1}{8\pi^2\rho\beta k}\int\limits_0^\infty
 \frac{p^2 \rd p}{\left[z_0^{-1}\exp(\beta p)-1\right]^2}
 \left(-\ln\left\{\frac{z_0^{-1}\exp\left[\beta (p+k)^2\right]-1}
 {z_0^{-1}\exp\left[\beta (p+k)^2\right]-1}\right\}\right.\nonumber\\
 &+4\beta pk -\left.\frac{1}{z_0^{-1}\exp\left[\beta (p+k)^2\right]-1}+\frac{1}{z_0^{-1}\exp\left[\beta (p-k)^2\right]-1}\right).
\end{align}
As a result we obtain:
\begin{align}\label{c2T}
c^2(T)\,=\;&\rho\nu_0+\frac{T}{m}\left\{S_2^0(T)+\frac{1}{2N}
\sum_{\mathbf{k}\neq0}\frac{\lambda_{k}S_4^0(k,T)}{1+\lambda_{k}S_0(k)}
-\frac{1}{2N}\sum_{\mathbf{k}\neq0}
\frac{\left[\lambda_k S_3^0(k,T)\right]^2}{[1+\lambda_{k}S_0(k)]^2}\right.
\nonumber\\
&+4C_2^0(T)+\frac{12}{N}\sum_{\mathbf{k}\neq0}\frac{C_3^0(k,T)S_3^0(k,T)}
{[1+\lambda_{k}S_0(k)]^2}
+\frac{8}{N}\sum_{\mathbf{k}\neq0}\frac{C_4^0(k,T)S_0(k)}{1+\lambda_{k}S_0(k)}
+\left.\frac{72}{N}\sum_{\mathbf{k}\neq0}
 \frac{\left[C_3^0(k,T)S_0(k)\right]^2}{[1+\lambda_{k}S_0(k)]^2}\right\}.
\end{align}

\section{Low and high temperature limit}

In the low temperature limit
\begin{align}
&\lim_{T\rightarrow 0}TC_2^0(T)=\frac{1}{32N}\sum_{{\bf k}\neq
0}\frac{\varepsilon_k\left(\alpha_{k}^2-1\right)^2}{\alpha_{k}^3},
& &\lim_{T\rightarrow 0}TC_3^0(k,T)=0, &  &\lim_{T\rightarrow
0}TC_4^0(k,T)=0,\nonumber\\
&\lim_{T\rightarrow 0}T S_2^0(k,T)=0,
& &\lim_{T\rightarrow 0}TS_3^0(k,T)=0, & & \lim_{T\rightarrow 0}TS_4^0(k,T)=0.
\end{align}
That is why
\begin{align}\label{c^2T0}
c^2\equiv\lim_{T\rightarrow
0}c^2(T)&=\frac{\rho\nu_0}{m}-\frac{1}{8mN}\sum_{{\bf k}\neq
0}\frac{\varepsilon_k\left(\alpha_{k}^2-1\right)^2}{\alpha_{k}^3}\,.
\end{align}
The same result we obtain if we take the second derivative of the energy (in the pair correlation approximation)
on the number of particles:
\begin{align}
c^2=\frac{N}{m}\frac{\partial^2 E}{\partial N^2}\,.
\end{align}

Taking the square root from the equation (\ref{c^2T0}) and using the smallness of the second term compared with the first one,
we have:
\begin{align}\label{cT0}
c\equiv\lim_{T\rightarrow
0}c(T)&=\sqrt{\frac{\rho\nu_0}{m}}-\frac{1}{16N\sqrt{m\rho\nu_0}}\sum_{{\bf
k}\neq
0}\frac{\varepsilon_k\left(\alpha_{k}^2-1\right)^2}{\alpha_{k}^3}\,.
\end{align}
The other way to the same result goes through formula \cite{VakUhn79,VakUhn80}:
\begin{align}
c=-\frac{\hbar}{m}\lim_{q\rightarrow0}q\tilde{a}_2(q),
\end{align}
where the value $\tilde{a}_2(q)$ has such a look:

\begin{align}
\tilde{a}_2({q})&=-\frac{1}{2}(\alpha_{q}-1)+\frac{1}{N}\sum_{{\bf
k}\neq0} \left[\frac{k^2}{2q^2\alpha_{q}} a_4{({\mathbf
q},-{\mathbf q},{\bf k},-{\bf
k})}\right.
+\left.\frac{({\bf k},{\mathbf q}+{\bf k})}
{q^2\alpha_{q}}a_3{({\mathbf q},{\bf k},-{\mathbf
q}-{\bf k})} \right].
\end{align}
In the high temperature limit, the contributions of three- and four-particle correlations are equal to zero \cite{VakHryh4}.
Taking into account that $\lim\limits_{T\rightarrow 0}S_0(q)=1$, we obtain a well-known classical expression for
the first sound velocity in a high temperature region \cite{Vak2004}:

\begin{eqnarray}
 c(T)=\sqrt{\frac{T}{m}+\frac{\rho\nu_0}{m}}\,.
\end{eqnarray}

\section{Numeric calculations}

We use the expression (\ref{c2T}) for the first sound velocity numeric calculation.
The transition from the summation to integration is carried out quite simply:
\begin{eqnarray}
\frac{1}{N}\sum_{{\bf k}\neq0}=\frac{1}{2\pi^2\rho}\int\limits_{0}^\infty k^2\rd k.
\end{eqnarray}
We can find the unknown quantity $\nu_0$ using the value of the first sound velocity at zero temperature,
which we obtain by extrapolating the  experimental data. Therefore, from relation
(\ref{c^2T0}) we have:
\begin{eqnarray}\label{nu_0}
\nu_0=\frac{m}{\rho}\left[c^2+\frac{1}{8mN}\sum_{k\neq
0}\frac{\varepsilon_k\left(\alpha_{k}^2-1\right)^2}{\alpha_{k}^3}\right].
\end{eqnarray}

We use the value for $\nu_0$ (\ref {nu_0}) only in the expressions that
reproduce the approximation of pair correlations in order not to exceed accuracy. We use $\nu_0=mc^2/\rho$
in expressions under the sum.
We do numeric calculation using the effective mass of helium atom in liquid instead of real mass,
in order to eliminate infra-red divergence \cite{HryhPryt}.
The use of the effective mass is a phenomenological one, but it is a necessary step, which allows us to get rid of
the essential consequences of approximate calculations (breaking of the series of the perturbation theory):
the infra-red divergences and the incorrect Bose-Einstein condensation temperature.

{We get the temperature behaviour of the isothermal sound velosity ($c_T$)  based on the experimental data for the adiabatic sound velocity ($c_\sigma$) \cite{DonBar}, specific heat at a constant volume ($C_v$) and
specific heat at a constant pressure ($C_p$) \cite{Arp} using the well-known relationship:  $c_T=c_{\sigma}\sqrt{{C_v}/{C_p}}$.}

\begin{figure}[htb]
\centerline{
\includegraphics[width=0.7\textwidth]{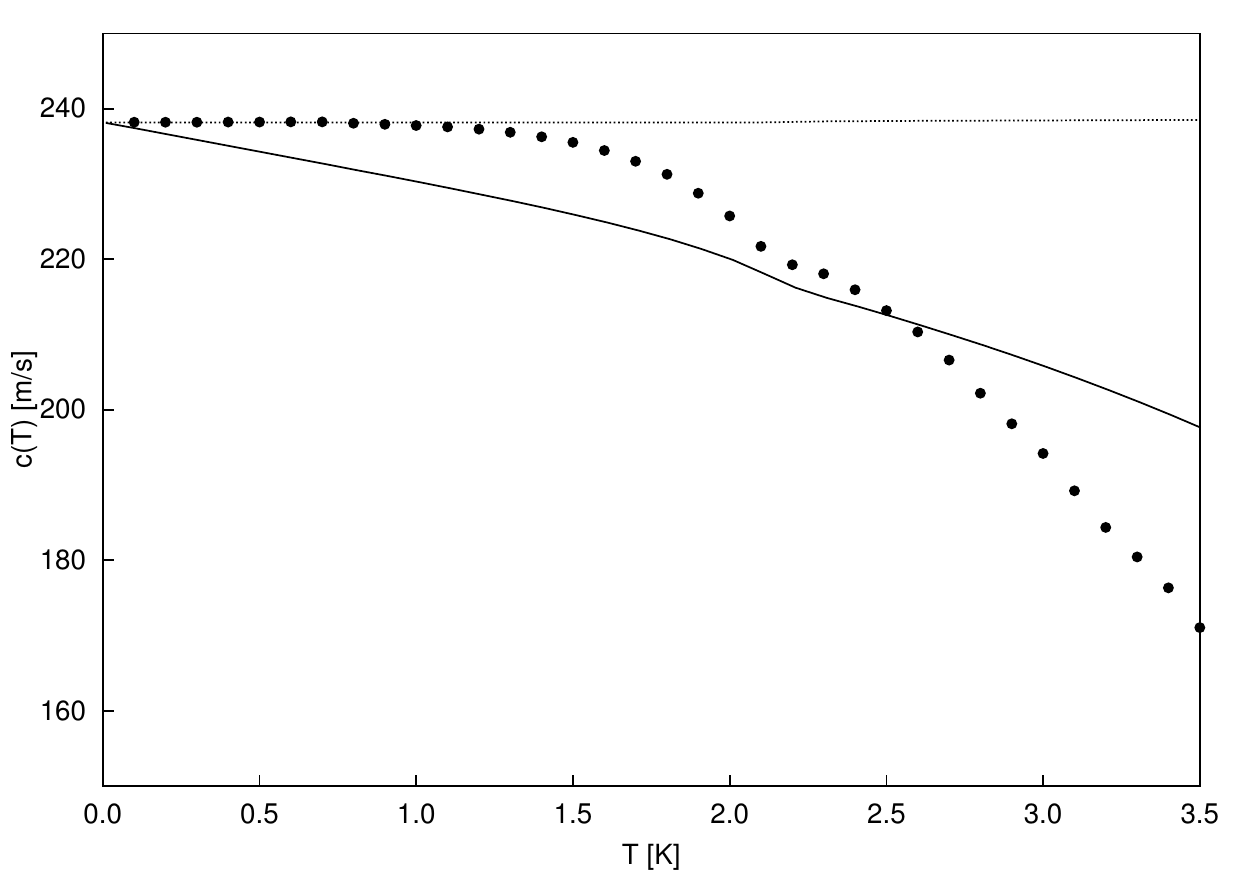}
}
\caption{Temperature dependence of the first sound velocity in liquid $^4$He.
Dashed curve~--- pair correlations approximation,
solid curve~--- post-RPA approximation, points~--- indirect experimental data ($c_T$).}
\label{fig1}
\end{figure}

\section{Conclusions}
In this paper we have got a temperature dependence of the first sound velocity in liquid $^4$He.
The received expression for the first sound velocity matches with the well-known results in both low and high
temperature limits. As it can be seen in figure~\ref{fig1}, the matching of the theoretical
results with the experimental data is quite good but not sufficient. In order to improve it, we should take into
account the next approximation for the sound velocity. It means that we need to use the expression for
a structure factor in ``two sums over the wave vector'' approximation in our calculations.

\section*{Acknowledgements}
I am greatful to my supervisor Prof.~I.O.~Vakarchuk for valuable remarks and suggestions, as well as to my colleague Dr.~V.S.~Pastukhov for longlasting debates and discussions concerning the topic of this paper.

\appendix

\section{Appendix}
Two-particle structure factor of the ideal bose gas \cite{VP2008}:
\begin{align}
 S_0(q)=1+2\frac{n_0}{N}n_q+\frac{1}{N}\sum\limits_{\bf{p}\neq0}n_pn_{|\bf{p}+\bf{q}|}\,.
\end{align}
Three-particle structure factor of the ideal bose gas:
\begin{align}
 S_0^{(3)}({\bf q},{\bf k},-{\bf q}-{\bf k})\,=\;&2\frac{n_0}{N}\left(n_q n_k+n_q n_{|{\bf q}+{\bf k}|}+n_k
 n_{|{\bf q}+{\bf k}|}\right)
 +S_0(q)+S_0(k)+S_0(|{\bf q}+{\bf k}|)-2\nonumber\\
 &+\frac{2}{N}\sum\limits_{{\bf p}\neq0} n_p n_{|{\bf p}+{\bf q}|}n_{|{\bf p}-{\bf k}|}\,.
\end{align}
Four-particle structure factor of the ideal bose gas after the elimination of the infra-red divergence:
\begin{align}
 S_0^{(4)}({\bf q},-{\bf q},{\bf k},-{\bf k})\,=\;&2\frac{n_0}{N}\left(n_{|{\bf q}-{\bf k}|}+n_{|{\bf q}+{\bf k}|}\right)
 n_k(1+n_k)\nonumber\\
 &+2\left[S_0^{(3)}({\bf q},{\bf k},-{\bf q}-{\bf k})-S_0(q)-S_0(k)+1\right]\nonumber\\
 &+\frac{2}{N}\sum\limits_{{\bf p}\neq0} n_p n_{|{\bf p}+{\bf q}|}n_{|{\bf p}+{\bf k}|}n_{|{\bf p}+{\bf q}+{\bf k}|}\,.
\end{align}

\ukrainianpart

\title{Швидкість першого звуку в рідкому $^4$He}
\author{О.І. Григорчак}
\address{ Кафедра теоретичної фізики, Львівський національний університет імені Івана Франка,\\
вул. Драгоманова, 12, 79005 Львів, Україна
}

\makeukrtitle

\begin{abstract}
\tolerance=3000%
На основі виразу для структурного фактора багатобозонної системи з урахуванням прямих три- і чотиричастинкових
кореляцій знайдено температурну поведінку швидкості першого звуку в рідкому $^4$He в пост-RPA наближенні.
У границі як низьких, так і високих температур отриманий вираз переходить у вже відомий.
Результати можуть бути застосовані для аналізу внесків три- та чотиричастинкових кореляцій у термодинамічні та структурні
функції рідкого $^4$He.
\keywords Бозе системи, рідкий $^4$He, швидкість першого звуку

\end{abstract}

\end{document}